\documentclass[12pt]{article}

\usepackage{latexsym}
\usepackage{amssymb,amsfonts,amsmath}
\usepackage{graphicx} 
\usepackage{indentfirst}
\usepackage{bbm}
\usepackage{amssymb}
\usepackage[normalem]{ulem}
\usepackage{verbatim}
\usepackage{amsmath, amsthm, amssymb}
\usepackage{mathrsfs}
\usepackage{hyperref}
\usepackage{amsfonts}
\usepackage{dsfont}
\usepackage{cite}
\usepackage{xcolor}
\usepackage{enumitem}  
\usepackage[open,openlevel=2,atend,numbered]{bookmark}

\parskip=\medskipamount

\parskip=\medskipamount

\newcommand {\cD}{{\cal D}}

\newcommand {\cN}{{\cal N}}

\newcommand {\cP}{{\cal P}}

\newcommand {\cR}{{\cal R}}

\newcommand {\cX}{{\cal X}}

\renewcommand{\a}{\alpha}
\renewcommand{\b}{\beta}
\renewcommand{\c}{\chi}
\renewcommand{\d}{\delta}
\newcommand{\e}{\epsilon}
\newcommand{\ce}{\varepsilon}
\newcommand{\f}{\phi}

\newcommand{\g}{\gamma}
\newcommand{\h}{\eta}

\renewcommand{\k}{\kappa}
\renewcommand{\l}{\lambda}
\newcommand{\m}{\mu}

\newcommand{\s}{\sigma}

\newcommand{\y}{\psi}

\newcommand{\G}{\Gamma}
\renewcommand{\L}{\Lambda}

\newcommand{\Q}{\Theta}

\newcommand{\W}{\Omega}

\newcommand{\Y}{\Psi}
\def\tr{{\rm Tr}}
\def\rd{{\rm d}}
\def\ri{{\rm i}}

\newcommand{\ii}{\mathrm{i}}
\def\intd4x{ \int \,\mathrm{d}^4 x \hspace{0.05cm} }

\newcommand{\ph}{\phantom}
\newcommand{\id}{\mathds{1}}

\newcommand{\ad}{{\dot{\alpha}}}                           
\newcommand{\bd}{{\dot{\beta}}}                            

\newcommand{\hf}{\frac12}

\newcommand{\be}{\begin{equation}}
	\newcommand{\ee}{\end{equation}}
\newcommand{\bea}{\begin{eqnarray}}
	\newcommand{\eea}{\end{eqnarray}}
\newcommand{\beal}{\begin{align}}
	\newcommand{\eeal}{\end{align}}
\newcommand{\bes}{\begin{subequations}}
	\newcommand{\ees}{\end{subequations}}

\newcommand{\dsR}{{\mathbb R}}

\newcommand{\dsA}{{\mathbb A}}
\newcommand{\dsT}{{\mathbb T}}

\def\double #1{#1{\hbox{\kern-2pt $#1$}}}

\newcommand{\hodge}{{\star}}

\newcommand{\mfS}{\mathfrak{S}}
\newcommand{\mfA}{\mathfrak{A}}
\newcommand{\mfF}{\mathfrak{F}}
\newcommand{\dd}{\mathrm{d}}

\newcommand{\sSL}{\mathsf{SL}}

\newcommand{\sSO}{\mathsf{SO}}
\newcommand{\sISO}{\mathsf{ISO}}
\newcommand{\sISL}{\mathsf{ISL}}

\newif\ifdtup

\newcommand{\bsubeq}{\begin{subequations}}
	\newcommand{\esubeq}{\end{subequations}}

\numberwithin{equation}{section}

\begin{document}

\begin{titlepage}	
\begin{flushright}
April, 2023\\
\end{flushright}
\vspace{5mm}

\begin{center}
{\Large \bf Nonlinear realisation approach to topologically massive supergravity}
\end{center}
	\nopagebreak

\begin{center}		
{\bf	Sergei M. Kuzenko and Jake C. Stirling}
		\vspace{5mm}
		
\footnotesize{
			{\it Department of Physics M013, The University of Western Australia\\
				35 Stirling Highway, Perth W.A. 6009, Australia}}
		\vspace{2mm}
		~\\
		Email: \texttt{sergei.kuzenko@uwa.edu.au, jake.stirling@research.uwa.edu.au}\\
		\vspace{2mm}

	\end{center}

\begin{abstract}
\baselineskip=14pt
We develop a nonlinear realisation approach to topologically massive supergravity in three dimensions, with and without a cosmological term. It is a natural generalisation of a similar construction for ${\cal N}=1$ supergravity in four dimensions, which was recently proposed by one of us. At the heart of both formulations is the nonlinear realisation approach to gravity which was given by Volkov and Soroka fifty years ago in the context of spontaneously broken local supersymmetry. In our setting, the action for cosmological topologically massive supergravity is invariant under two different local supersymmetries. One of them acts on the Goldstino, while the other supersymmetry leaves the Goldstino invariant. The former can be used to gauge away the Goldstino, and then the resulting action coincides with that given in the literature.

\end{abstract}
\vspace{5mm}

\begin{flushright}
{\it Dedicated to the memory of Stanley Deser}
\end{flushright}

\vfill

\vfill
\end{titlepage}

\newpage
\renewcommand{\thefootnote}{\arabic{footnote}}
\setcounter{footnote}{0}

	\tableofcontents{}
	\vspace{1cm}
	\bigskip\hrule

\allowdisplaybreaks

\section{Introduction}

The method of nonlinear realisations of groups (also known as the coset construction), 
which was systematically developed by Coleman, Wess and Zumino
\cite{CWZ, CCWZ} (see also \cite{Isham, SalamS1}), 
is the mathematical formalism to construct phenomenological Lagrangians describing the low-energy dynamics of Goldstone fields in theories with spontaneously broken symmetry. This method was extended to spacetime symmetries by Volkov \cite{Volkov} (see also \cite{Ogievetsky}), although the case of spontaneously broken conformal symmetry had been studied earlier \cite{SalamS2,ISS1,ISS2,Zumino}.
In modern applications of the method of nonlinear realisations, an important role is played by the inverse Higgs mechanism discovered by Ivanov and Ogievetsky \cite{IO}. An interesting interpretation of this mechanism was given in \cite{McArthur}.

The formalism of nonlinear realisations can also be used to construct gauge theories, including those describing gravity and its matter couplings. The importance of nonlinear realisations for gravity was realised fifty years ago by Volkov and Soroka \cite{VS, VS2} (for related  
developments see \cite{SW2, IvanovN}).\footnote{Gauge formulations for general relativity have been discussed since the pioneering work by Utiyama \cite{Utiyama}, Kibble \cite{Kibble} and Sciama \cite{Sciama}, see e.g.\cite{Hehl, Blagojevic:2013xpa} for reviews. } 
These authors gauged the $\cN$-extended super-Poincar\'e group in four dimensions (4D) and proposed a super-Higgs mechanism by constructing the $\cN=1$ supergravity action with nonlinearly realised local supersymmetry
(see \cite{Volkov2} for a review and \cite{BMST} for a critical analysis of the Volkov-Soroka construction and modern developments).
Restricting their analysis to the $\cN=0$ case results in the nonlinear realisation approach to gravity, which corresponds to the coset space $\sISL(2,{\mathbb C}) / \sSL(2,{\mathbb C}) $
in the formulation of \cite{VS2}.\footnote{The group  $\sISL(2,{\mathbb C}) $ is isomorphic to the 
universal covering group, $\widetilde{\sISO}_0(3,1)$, of the proper orthochronous Poincar\'e group
$\sISO_0(3,1)$.}
A review of this construction is given in appendix \ref{Appendix0}.
The theory is described by a vielbein, an independent Lorentz connection and a vector Goldstone 
field $V^a$. There are  two types of gauge transformations with vector-like parameters, the general coordinate transformations and the local Poincar\'e translations. The latter gauge freedom acts on the Goldstone field by the rule $V'^a = V^a +b^a$ and, therefore, 
it can be fixed by imposing the condition $V^a =0$. As a result, one arrives at the first-order formulation for gravity \cite{Kibble}. The Goldstone field in this setting is a compensator. In the terminology of \cite{McArthur}, $V^a$ is an unphysical Goldstone boson describing purely gauge degrees of freedom.

Within the Volkov-Soroka approach to spontaneously broken local supersymmetry \cite{VS, VS2}, there are two Goldstone fields, the vector field $V^a$ and a spinor field $(\psi^\a , \bar \psi_\ad )$. The latter is 
the Goldstone field for supersymmetry transformations\footnote{Here we restrict our discussion to the $\cN=1$ case.} \cite{VA,AV}. It is called the Goldstino. While $V^a$ is an unphysical Goldstone boson, the Goldstino is in general a genuine Goldstone field for it triggers spontaneous breakdown of the local supersymmetry. In the gauge $\psi^\a =0 $, the gravitino becomes massive. 
 A natural question is the following.  Is it possible to have a dynamical system such that $\psi^\a$ turns into an unphysical Goldstone field? The positive answer was given in \cite{K2021} where it was shown  that, for specially chosen parameters of the theory, the Volkov-Soroka action is invariant under two different local supersymmetries. One of them is present for arbitrary values of the parameters and acts on the Goldstino, while the other supersymmetry emerges only in a special case and leaves the Goldstino invariant.  The former can be used to gauge away the Goldstino, and then the resulting action coincides with that proposed by Deser and Zumino for consistent supergravity in the first-order formalism 
 \cite{DZ}.\footnote{The $\cN=1$ supergravity in the second-order formalism \cite{FvNF} is obtained by using the equations of motion for the Lorentz connection to express it in terms of the other fields.}

In this paper we will extend the construction of \cite{K2021} to the case of 3D $\cN=1$ supergravity \cite{HT,HT2}, with and without a cosmological term, and then the obtained results will be generalised to topologically massive $\cN=1$ supergravity \cite{DeserKay} and its cosmological extension \cite{Deser}.

It should be pointed out that the literature on simple supergravity in three dimensions is immense. In particular, superfield approaches to $\cN=1$ supergravity-matter systems were developed, e.g., in \cite{BG,GGRS,ZP88,ZP89, LR89, KLT-M11}.  
The $\cN=1$ supersymmetric Lorentz Chern-Simons term \cite{DeserKay}, which is at the heart of (cosmological) topologically massive supergravity \cite{DeserKay,Deser}, has been interpreted as the action for 3D $\cN=1$ conformal supergravity \cite{vN}.\footnote{The structure of 3D $\cN=1$ conformal supergravity was also studied in \cite{Uematsu}.} Superfield formulations for $\cN=1$ conformal supergravity were derived in \cite{KT-M12,BKNT-M1,BKNT-M2} (somewhat incomplete results had appeared earlier in \cite{GGRS,ZP88,ZP89}). The Chern-Simons formulation for $\cN=1$ anti-de Sitter (AdS) supergravity was proposed in \cite{AT}.
The super-Higgs effect for $\cN=1$ supergravity was first described in 
\cite{Dereli:1977yx}. The Hamiltonian form of (topologically) massive $\cN=1$ supergravity was constructed in \cite{Hohm:2012vh,Routh}.

This paper is organised as follows. In section \ref{sect2} we present a 3D analogue of the Volkov-Soroka construction. Using this framework, we demonstrate in section \ref{sect3} that the action for pure $\cN=1$ Poincar\'e supergravity \eqref{3DSUGRA} is invariant under two different local supersymmetries. One of them 
is present for an arbitrary relative coefficient 
between the two terms in \eqref{3DSUGRA} 
 and acts on the Goldstino, while the other supersymmetry emerges only in a special case and leaves the Goldstino invariant. The former can be used to gauge away the Goldstino, and then the resulting action coincides with the standard action for Poincar\'e supergravity in the first-order formalism. In subsection \ref{subsect3.2}
 we show that the same formalism of nonlinearly realised local supersymmetry can be used to describe AdS supergravity, however the second local supersymmetry has to be deformed.  In section \ref{sect4} we generalise the analysis of section \ref{sect3} to topologically massive supergravity and its cosmological extension. The main body of the paper is accompanied by three technical appendices. 
 Appendix \ref{Appendix0} reviews the nonlinear realisation approach to 4D gravity.
 In appendix \ref{Appendix A} we collect the key formulae of the 3D two-component spinor formalism. Finally, appendix \ref{appendixB} derives
the first Bianchi identity.


\section{The Volkov-Soroka approach in three dimensions}\label{sect2}
	
Let $\cP(3|\cN)$ be the three-dimensional $\cN$-extended super-Poincar\'e group. Any element $g\in\cP(3|\cN)$ is a $(4|\cN)\times(4|\cN)$ supermatrix of the form{\footnote{Our parametrisation of the elements of $\cP(3|\cN)$ follows \cite{KPT-MvU}.}}
	\begin{subequations}
		\begin{align}
			g&= g(b,\h,M,\cR) = s(b,\h)h(M,\cR)\equiv sh~,\\
			s(b,\h)&:=\left(
			\begin{array}{c|c|c}
				\id_2 & 0 & 0 \\
				\hline
				-\hat{b}+\frac{\ii}{2}\ce^{-1}\h^2& \id_2 & -\sqrt{2}\h^T \\
				\hline
				\ii\sqrt{2}\h & 0 & \id_{\cN}
			\end{array}
			\right) =\left(
			\begin{array}{c|c|c}
				\d_\a\,^\b & 0 & 0 \\
				\hline
				-b^{\a\b}+\frac{\ii}{2}\ce^{\a\b}\h^2& \d^\a\,_\b & -\sqrt{2}\h^\a\,_J \\
				\hline
				\ii\sqrt{2}\h_I\,^\b & 0 & \d_{IJ}
			\end{array}
			\right)~,\\
			h(M,\cR)&:= \left(
			\begin{array}{c|c|c}
				M & 0 & 0 \\
				\hline
				0 & (M^{-1})^T & 0 \\
				\hline
				0 & 0 & \cR
			\end{array}
			\right) =\left(
			\begin{array}{c|c|c}
				M_\a\,^\b & 0 & 0 \\
				\hline
				0 & (M^{-1})_\b\,^\a & 0 \\
				\hline
				0 & 0 & \cR_{IJ}
			\end{array}
			\right)~,
		\end{align}
	\end{subequations}
	where $M\in \sSL(2,\dsR)$, $\cR\in \sSO(\cN)$, 
	$\eta = (\eta_I{}^\b)$, 
	 $\h^2:=\h_I^\a\h_{\a I}$, and $\hat b$ is defined in \eqref{B3.b}.
 The $\sSL(2,\dsR)$ invariant spinor metric $\ce=(\ce_{\a\b}) = -(\ce_{\b \a}) $ and its inverse $\ce^{-1}=(\ce^{\a\b}) = - (\ce^{\b\a}) $ are
 defined in appendix \ref{Appendix A}. 
 The group element $s(b,\h)$ is labelled by three bosonic real parameters $b^a$ and $2\cN$ fermionic real parameters $\h_I\,^\a=\h^\a\,_I\equiv\h_I^\a$. 
	
Let us introduce Goldstone fields $Z^A(x)=\left(X^a(x),\Q_I^\a(x)\right)$ for spacetime translations $(X^a)$ and supersymmetry transformations $(\Q_I^\a)$. They parametrise the homogeneous space ($\cN$-extended Minkowski superspace)
	\begin{equation}
		\mathbb{M}^{3|2\cN}=\frac{\cP(3|\cN)}{\sSL(2,\dsR)\times\sSO(\cN)}
	\end{equation}
	according to the rule:
	\begin{equation}
		\mfS(Z)=\left(
		\begin{array}{c|c|c}
			\id_2 & 0 & 0 \\
			\hline
			-\hat{X}+\frac{\ii}{2}\ce^{-1}\Q^2& \id_2 & -\sqrt{2}\Q^T \\
			\hline
			\ii\sqrt{2}\Q & 0 & \id_{\cN}
		\end{array}
		\right)\Longrightarrow \mfS^{-1}(Z)=\left(
		\begin{array}{c|c|c}
			\id_2 & 0 & 0 \\
			\hline
			\hat{X}+\frac{\ii}{2}\ce^{-1}\Q^2& \id_2 & \sqrt{2}\Q^T \\
			\hline
			-\ii\sqrt{2}\Q & 0 & \id_{\cN}
		\end{array}
		\right).
	\end{equation}
	
	A gauge super-Poincar\'e transformation acts as 
	\begin{equation}
		g(x):Z(x)\rightarrow Z'(x)~,\qquad g\mfS(Z)=\mfS(Z')h~,
	\end{equation}
	with $g=sh$. This is equivalent to the following transformations of the Goldstone fields:
	\begin{subequations}
		\begin{align}\label{stransGoldstone}
			s(b,\h):\qquad \hat{X}'&=\hat{X}+\hat{b}+\ii(\h^{\rm{T}}\Q-\Q^{\rm{T}}\h)~,\\
			\Q'&=\Q+\h~\label{stransGoldstino},
		\end{align}
	\end{subequations}
	and
	\begin{subequations}
		\begin{align}
			h(M,\cR):\qquad \hat{X}'&=(M^{-1})^{\rm{T}}\hat{X}M^{-1}~,\\
			\Q'&=\cR\Q M^{-1}~.
		\end{align}
	\end{subequations}
	
	Introduce a connection $\mfA=\dd x^m\mfA_m$ taking its values in the super-Poincar\'e algebra,	
	\begin{equation}
		\mfA:=\left(
		\begin{array}{c|c|c}
			\hf \W & 0 & 0 \\
			\hline
			-\hat{e}& - \hf\W^{\rm{T}} & -\sqrt{2}\y^{\rm{T}} \\
			\hline
			\ii\sqrt{2}\y & 0 & r
		\end{array}
		\right)=\left(
		\begin{array}{c|c|c}
			\hf \W_\a\,^\b & 0 & 0 \\
			\hline
			-e^{\a\b}& - \hf \W^\a\,_\b & -\sqrt{2}\y^\a\,_J \\
			\hline
			\ii\sqrt{2}\y_I\,^\b & 0 & r_{IJ}
		\end{array}
		\right),
	\end{equation}
	and possessing the gauge transformation law
	\begin{equation}
		\mfA'=g\mfA g^{-1}+g\dd g^{-1}.
	\end{equation}
	Here the one-form $\W_\a\,^\b$ is related to the Lorentz connection $\W^{ab}=\dd x^m\W_m\,^{ab}=-\W^{ba}$ as
	\begin{equation}
		\W_\a\,^\b=\frac{1}{2}\ce_{abc}(\g^a)_\a\,^\b\W^{bc}.
	\end{equation}
	As in the first-order formalism to gravity, the Lorentz connection is an independent field and may be expressed in terms of the other fields by requiring it to be on-shell. The one-form $e^{\a\b}$ is the spinor counterpart of the dreibein $e^a=\dd x^m e_m\,^a$. The fermionic one-forms $\y_I\,^\b$ describe $\cN$ gravitini. Finally, the one-form $r_{IJ} = -r_{JI}$ is the $\sSO(\cN)$ gauge field.
	
It should be pointed out that our parametrisation of the super-Poincar\'e algebra follows \cite{BKNT-M2} and differs from \cite{KPT-MvU}. Under an infinitesimal Lorentz transformation 
\begin{subequations}
\bea
\d x^a = \l^a{}_b x^b = \varepsilon^{abc} \l_b x_c~, \qquad \l_{ab} = -\l_{ba}
\eea
a two-component spinor $\psi_\a$ transforms as 
\bea
\d \psi_\a = \hf \l_\a{}^\b \psi_\b~, \qquad \l_{\a\b} = \l_{\b\a}~,
\eea 
\end{subequations}
where the Lorentz parameters $\l_{ab}$, $\l_a$ and $\l_{\a\b}$ are related to each other according to the rules \eqref{B.11}, \eqref{B.12} and \eqref{B.13}.

	Associated with $\mfS$ and $\mfA$ is the different connection
	\begin{equation}
		\dsA:=\mfS^{-1}\mfA\mfS+\mfS^{-1}\dd\mfS~,
	\end{equation}
	with gauge transformation law
	\begin{equation}\label{dsAtrans}
		\dsA'=h\dsA h^{-1}+h\dd h^{-1}~,
	\end{equation}
	for an arbitrary gauge parameter $g=sh$. This connection is the main object in the Volkov-Soroka construction. 
	Direct calculations give the explicit form of $\dsA$
	\begin{equation}
		\dsA:=\left(
		\begin{array}{c|c|c}
			\hf \W & 0 & 0 \\
			\hline
			-\hat{E} & -\hf \W^{\rm{T}} & -\sqrt{2}\Y^{\rm{T}} \\
			\hline
			\ii\sqrt{2}\Y & 0 & r
		\end{array}
		\right),
	\end{equation}
	where we have defined 
	\begin{subequations}
		\begin{align}
			\hat{E}&:=\hat{e}+\cD\hat{X}+\ii\left(\cD\Q^{\rm{T}}\Q-\Q^{\rm{T}}\cD\Q\right)+2\ii\left(\y^{\rm{T}}\Q-\Q^{\rm{T}}\y\right)~,\\
			\Y&:=\y+\cD\Q~,\qquad
			\Y^{\rm{T}}=\y^{\rm{T}}+\cD\Q^{\rm{T}}~,
		\end{align}
	\end{subequations}
	and $\cD$ denotes the covariant derivative,
	\begin{subequations}
		\begin{align}
			\cD\hat{X}&=\dd\hat{X}- \hf \hat{X}\W- \hf \W^{\rm{T}}\hat{X}~,\\
			\cD\Q&=\dd\Q-\hf \Q\W+r\Q~,\qquad
			\cD\Q^{\rm{T}}=\dd\Q^{\rm{T}}- \hf\W^{\rm{T}}\Q^{\rm{T}}-\Q^{\rm{T}}r ~.
		\end{align}
	\end{subequations}
	Equation (\ref{dsAtrans}) is equivalent to the following gauge transformation laws:
	\begin{subequations}
		\begin{align}
			\W'&=M\W M^{-1}+M\dd M^{-1}~,\\
			r'&=\cR r\cR^{-1}+\cR\dd\cR^{-1}
		\end{align}
	\end{subequations}
	and
	\begin{subequations}
		\begin{align}
			\hat{E}'&=(M^{-1})^{\rm{T}}\hat{E}M^{-1}~,\\
			\Y'&=\cR\Y M^{-1}~.
		\end{align}
	\end{subequations}
	It is worth pointing out that the supersymmetric one-forms $E^a$ and $\Y_I\,^\b$ transform as tensors with respect to the Lorentz and $\sSO(\cN)$ gauge groups.
	
	Under a supersymmetry transformation, $g=s(0,\h)$, one can use the Goldstone field transformations (\ref{stransGoldstone}) and (\ref{stransGoldstino}) to deduce the local supersymmetry transformation laws of the gravitini and the dreibein
	\begin{subequations}\label{SUSYelem}
		\begin{align}\label{SUSYelem1}
			\y'&=\y-\cD\h~,\\
			\hat{e}'&=\hat{e}+2\ii\left(\h^{\rm{T}}\y-\y^{\rm{T}}\h\right)+\ii\left(\cD\h^{\rm{T}}\h-\h^{\rm{T}}\cD\h\right)\label{SUSYelem2}~.
		\end{align}
	\end{subequations}
	In the infinitesimal case, these supersymmetry transformation laws take the form
	\begin{subequations}\label{2.188}
	\begin{equation}
		\d_\h\y=-\cD\h~,\qquad\d_\h e^a=2\ii\,{\rm tr}(\h\g^a\y^{\rm{T}})~.
	\end{equation}
	These should be accompanied by the supersymmetry transformations of the Goldstone fields 
\bea
\d_\h X^a = -\ri {\rm tr} (\Theta \g^a \eta^{\rm T})~, \qquad \d_\h \Theta = \eta~.
\eea
\end{subequations}
	
A local Poincar\'e translation is given by $g=s(b,0)$. It acts on the Goldstone vector field $X^a$ and the dreibein $e^a$ as follows
	\begin{equation}\label{transgauge}
		X'^a=X^a+b^a~,\qquad e'^a=e^a-\cD b^a~,
	\end{equation}
	while leaving the Goldstini and gravitini inert.
	
	The curvature tensor is found through 
	\begin{equation}
		\dsR=\dd\dsA-\dsA\wedge\dsA~,\qquad \dsR'=h\dsR h^{-1}~.
	\end{equation}
	Direct calculations give
	\begin{equation}
		\dsR:=\left(
		\begin{array}{c|c|c}
			\hf R & 0 & 0 \\
			\hline
			-\hat{\mathbb T}& -\hf R^{\rm{T}} & -\sqrt{2}\cD\Y^{\rm{T}} \\
			\hline
			\ii\sqrt{2}\cD\Y & 0 & F
		\end{array}
		\right)~,
	\end{equation}
	where $R=(R_\a\,^\b)$ is the Lorentz curvature, $F=(F_{IJ})$ is the Yang-Mills field strength,
		\begin{align}
			\cD\Y&=\dd\Y- \hf \Y\wedge\W-r\wedge\Y~,\qquad
			\cD\Y^{\rm{T}}=\dd\Y^{\rm{T}}+ \hf \W^{\rm{T}}\wedge\Y^{\rm{T}}-\Y^{\rm{T}}\wedge r
		\end{align}
	are the gravitino field strengths, and 
	\begin{equation}
		\hat{\mathbb T}=\dd\hat{E}- \hf\hat{E}\wedge\W+ \hf \W^{\rm{T}}\wedge\hat{E}-2\ii\Y^{\rm{T}}\wedge\Y=\cD\hat{E}-2\ii\Y^{\rm{T}}\wedge\Y
	\end{equation}
	is the supersymmetric torsion tensor. In vector notation, the torsion tensor reads
	\begin{equation}
		\dsT^a=\cD E^a-\ii\Y\wedge\g^a\Y^{\rm{T}}~.
	\end{equation}
	The Lorentz curvature tensor with spinor ($R_\a{}^\b$) and vector ($R^a{}_b)$ indices has the form 
\bea
R_\a{}^\b = \rd \Omega_\a{}^\b - \hf \W_\a{}^\g \wedge \W_\g{}^\b ~, \qquad 
R^a{}_b = \rd \Omega^a{}_b -  \W^a{}_c \wedge \W^c{}_b ~.
\eea

	Using the above results, one can construct a locally supersymmetric action. 
	With the notation $E= \det (E_m{}^a)$,
	gauge-invariant functionals include the following:
	\begin{itemize}
		\item The Einstein-Hilbert action
		\begin{equation}\label{EHaction}
			S_{\text{EH}}=\frac{1}{2}\int\ce_{abc}E^a\wedge R^{bc} =\frac{1}{2}\int \rd^3 x \, E\, R~;
		\end{equation}
		\item The Rarita-Schwinger action 
		\begin{equation}\label{RSaction}
			S_{\text{RS}}=\ii\int\Y_I^\a\wedge\cD\Y_{\a I} =\ii\int \rd^3 x \, E\, \ce^{mnp}\Y_{m\,I}^{\phantom{m\,}\a}\cD_p\Y_{n\a I}~;
		\end{equation}
		\item The cosmological term
		\begin{equation}
			S_{\text{cosm}}=-\frac{1}{6}\int\ce_{abc}E^a\wedge E^b\wedge E^c
		= \int \rd^3 x\, E	~;
		\end{equation}
		\item The mass term
		\begin{equation}
			S_{\text{mass}}=\int\Y_I\wedge E^a\g_a\wedge\Y_I =\int \rd^3 x \, E\, \ce^{mnp}\Y_{mI}\g_n\Y_{pI}~. \label{mass}
		\end{equation}
	\end{itemize}
	In contrast to the 4D case, the mass term is invariant under the entire $R$-symmetry group $\sSO(\cN)$. 
	Making use of the $\sSO(\cN) $ connection $r$ and the corresponding field strength $F$, we can construct standard Chern-Simons and Yang-Mills actions. We will not use them. In the $\cN=1$ case, a  linear combination of the above functionals gives an action for spontaneously broken supergravity.

\section{Second local supersymmetry}\label{sect3}
	
In the remainder of this paper our discussion is restricted to the $\cN=1$ case for simplicity. If $\cN>1$, it is necessary to take into  account the $\sSO(\cN)$ connection. An extension of our approach to the $\cN=2$ case will be studied elsewhere. 
	
	\subsection{Poincar\'e supergravity} 
	
	Each of the functionals \eqref{EHaction}--\eqref{mass} is invariant under the
	local supersymmetry transformation \eqref{2.188}.
We are going to show that a special linear combination of the actions 
\eqref{EHaction} and \eqref{RSaction} possesses a second local supersymmetry
described by  the parameter $\e=(\e^\a)$. This combination is 
	\begin{equation}\label{3DSUGRA}
		S_{\text{SG}}=S_{\text{EH}}-2S_{\text{RS}}=\frac{1}{2}\int\ce_{abc}E^a\wedge R^{bc}-2\ii\int\Y\wedge\cD\Y~.
	\end{equation}
Making use of the first supersymmetry transformation \eqref{2.188} and the local Poincar\'e translation \eqref{transgauge} allows us to impose the unitary gauge 
\bea
X^a =0~, \qquad \Theta^\a =0~.
\label{unitary}
\eea
Then \eqref{3DSUGRA} turns into the action for pure $\cN=1$ supergravity without a cosmological term \cite{HT}.

	Under the second supersymmetry,  the composite fields $E^a$ and $\Y^\a$ are postulated to transform as
	\begin{subequations} \label{2.222}
		\begin{equation}\label{comptrans}
			\d_\e\Y^\a=-\cD\e^\a,\qquad \d_\e E^a=2\ii\e\g^a\Y~.
		\end{equation}
		The Goldstone fields are required to be  inert under this transformation,
		\begin{equation}\label{connvar}
			\d_\e X^a=0~, \qquad \d_\e\Q^\a=0~.
		\end{equation}
		The elementary fields $\y^\a$ and $e^a$ transform as follows:
		\begin{align}\label{elemtrans}
			\d_\e\y^\a&=-\cD\e^\a+\hf (\Q\d_\e\W)^\a~,\\
			\d_\e e^a&=-\d_\e\W^a\,_bX^b+2\ii\e\g^a\Y+2\ii\cD\e\g^a\Q-\frac{\ii}{4}\ce^{abc}\d_\e\W_{bc}\Q^2~.
			\label{elemtrans2}
		\end{align}
	\end{subequations}
	The dependence on $\d_\e\W$ in (\ref{elemtrans}) and (\ref{elemtrans2}) is such that the composite fields $\Y^\a$ and $E^a$ remain unchanged when the connection gets the displacement $\W\rightarrow\W+\d_\e\W$. As will be shown, the 
	transformation law of $\Omega$ will be determined by demanding the action (\ref{3DSUGRA}) to be invariant under this new local supersymmetry 
	\eqref{2.222}.
	
	We now compute variations of the two terms in the action (\ref{3DSUGRA}). Denote $\d_\e^{(1)}$ for variations with respect to the transformations (\ref{comptrans}) and $\d_\e^{(2)}$ for variations with respect to the Lorentz connection. Computing the $\d_\e^{(1)}$ variation of the Einstein-Hilbert action (\ref{EHaction}) gives
	\begin{equation}\label{EHvar1}
		\d_\e^{(1)}S_{\text{EH}}=\ii\int\ce_{abc}R^{ab}\wedge\e\g^c\Y~.
	\end{equation}
	Computing the $\d_\e^{(1)}$ variation of the Rarita-Schwinger action (\ref{RSaction}) gives
	\begin{equation}\label{RSvar1}
		\d_\e^{(1)}S_{\text{RS}}=\frac{\ii}{2} \int\ce_{abc}R^{ab}\wedge\e\g^c\Y~,
	\end{equation}
	where we have used the relations
	\begin{equation}
		\cD\cD\Y=-\hf \Y\wedge R~,\qquad \cD\cD\e=-\hf \e R~.
		\label{3.66}
	\end{equation}
	As a result, computing the $\d_\e^{(1)}$ variation of the action (\ref{3DSUGRA}), we observe that the curvature contributions (\ref{EHvar1}) and (\ref{RSvar1}) precisely cancel each other,
	\begin{align}\label{SUGRAvar1}
		\d_\e^{(1)}S_{\text{SG}}&=\d_\e^{(1)}(S_{\text{EH}}-2S_{\text{RS}})
		=\d_\e^{(1)}S_{\text{EH}}-2\d_\e^{(1)}S_{\text{RS}}
		=0~.
	\end{align}
	
	Next, we vary the action (\ref{3DSUGRA}) with respect to the Lorentz connection $\W^{ab}$. We give the Lorentz connection a small disturbance $\W\rightarrow\W+\d_\e\W$, with $\d_\e\W$ to be determined below, and assume that the elementary fields $\y^\a$ and $e^a$ also acquire $\d_\e\W$-dependent variations given in (\ref{elemtrans}) and (\ref{elemtrans2}). For the Einstein-Hilbert action we get the variation
	\begin{equation}
		\d_\e^{(2)}S_{\text{EH}}=\hf \int\ce_{abc}\cD E^a\wedge\d_\e\W^{bc}~.
	\end{equation}
	The Rarita-Schwinger action variation is
	\begin{equation}
		\d_\e^{(2)}S_{\text{RS}}=\frac{\ii}{4}\int\ce_{abc}\Y\wedge\g^a\Y\wedge\d_\e\W^{bc}~.
	\end{equation}
	Hence, the variation of the total action (\ref{3DSUGRA}) with respect to the Lorentz connection is
	\begin{equation}\label{SUGRAvar2}
		\d_\e^{(2)}S_{\text{SG}}=\hf \int\dsT^a\wedge\ce_{abc}\d_\e\W^{bc}~.
	\end{equation}
	Combining the results (\ref{SUGRAvar1}) and (\ref{SUGRAvar2}), we end up with 
	\begin{equation}\label{3DSUGRAvar}
		\d_\e S_{\text{SG}}=\hf \int\dsT^a\wedge\ce_{abc}\d_\e\W^{bc}~.
	\end{equation}
	This variation vanishes if $\d_\e\W^{bc}=0$, which differs from the case of $\cN=1$ supergravity in four dimensions considered in \cite{K2021}.

Alternatively, we can work with a composite connection obtained by imposing the constraint 
	\begin{equation}\label{constraint}
		\dsT^a=\cD E^a-\ii\Y\wedge\g^a\Y=\dd E^a+E^b\wedge\W^a\,_b-\ii\Y\wedge\g^a\Y=0~.
	\end{equation}
In the case of vanishing Goldstone fields, $X^a=0$ and $\Q^\a=0$, one can uniquely solve (\ref{constraint}) for the connection giving its well-known expression in terms of the dreibein and gravitino, $\W=\W(e,\y)$.

It is a simple observation that \eqref{constraint} is the equation of motion for the Lorentz connection $\Omega$. If this equation holds, the explicit form of the variation $\d_\e\W$ is irrelevant when computing $\d_\e S_{\text{SG}}$.
 Thus the Volkov-Soroka approach allows one to naturally arrive at the 1.5 formalism \cite{TvN, CW}.

\subsection{Anti-de Sitter supergravity} \label{subsect3.2}
	
In order to describe a supersymmetric extension of gravity with a cosmological term
\bea
S= \hf \int \rd^3 x \, e\, ( R -2\L) ~,
\eea
the second supersymmetry transformation \eqref{2.222} has to be deformed.

	Let us alter the $\Y^\a$ transformation (\ref{comptrans}) in the following way
	\begin{equation}\label{comptranscosmo}
		\d_\e\Y^\a=-\cD\e^\a-\frac{m}{2}(\e\g_a)^\a E^a\equiv \d_\e^{(1)}\Y^\a+\d_\e^{(m)}\Y^\a~,
	\end{equation}
	while keeping the $E^a$ and Goldstone field transformations the same, as given by the equations  (\ref{comptrans}) and (\ref{connvar}). Here $m$ is a constant real  parameter. The elementary fields $\y^\a$ and $e^a$ pick up an additional term proportional to $m$:
	\begin{subequations}
		\begin{align}\label{elemtranscosmo}
			\d_\e\y^\a&=-\cD\e^\a+\hf (\Q\d_\e\W)^\a-\frac{m}{2}(\e\g_a)^\a E^a~,\\
			\d_\e e^a&=-\d_\e\W^{ab}X_b+2\ii\e\g^a\Y+2\ii\cD\e\g^a\Q-\frac{\ii}{4}\ce^{abc}\d_\e\W_{bc}\Q^2+\ii m(\e\g_b\g^a\Q)E^b~.
			\label{elemtranscosmo2}
		\end{align}
	\end{subequations}

		Let us now add to the action (\ref{3DSUGRA}) 
		a supersymmetric cosmological term
	\begin{align}\label{cosmoaction}
		S_{\text{super-cosm}}&=m^2S_{\text{cosm}}-\ii mS_{\text{mass}}\nonumber\\
		&=-\frac{1}{6}m^2\int\ce_{abc}E^a\wedge E^b\wedge E^c-\ii m\int\Y\wedge E^a \wedge\g_a\Y~.
	\end{align}
	We will show that the resulting additional variation for the action (\ref{3DSUGRA}) due to the term proportional to $m$ in (\ref{comptranscosmo}) combined with the total variation of the action (\ref{cosmoaction}) does not contribute to the already established variation (\ref{3DSUGRAvar}) if we require certain conditions.
	
	First we compute the additional variation of the action (\ref{3DSUGRA}),
	\begin{align}\label{additionalvariation}
		\d_\e^{(m)}S_{\text{SG}}&=-2\d_\e^{(m)}S_{\text{RS}}\nonumber\\
		&=-2\ii m\int\Big(\cD\e^\b(\g_a)_\b\,^\a\wedge E^a\wedge\Y_\a-(\e\g_a)^\a\cD E^a\wedge\Y_\a\Big)~,
	\end{align}
	where we have denoted $\d_\e^{(m)}$ for the variation due to the additional term $-\hf m(\e\g_a)^\a E^a$ appearing in (\ref{comptranscosmo}). The total variation of the action (\ref{cosmoaction}) under the transformations (\ref{comptrans}) and (\ref{comptranscosmo}) respectively reads{\footnote{There is no connection variation contribution from this action.}}
	\begin{align}\label{cosmovariation}
		\d_\e S_{\text{super-cosm}}&= 2\ii m\int\cD\e^\b(\g_a)_\b\,^\a\wedge E^a\wedge\Y_\a~.
	\end{align}
	Combining all variations (\ref{3DSUGRAvar}), (\ref{additionalvariation}) and (\ref{cosmovariation}), we end up with 
	\begin{align}
		\d_\e S_{\rm AdS} &=\hf \int\dsT^a\wedge\ce_{abc}\d_\e\W^{bc}+2\ii m\int\cD E^a\wedge\e\g_a\Y\nonumber\\
		&=\hf \int\dsT^a\wedge\ce_{abc}\d_\e\W^{bc}+2\ii m\int(\dsT^a+\ii\Y\wedge\g^a\Y)\wedge\e\g_a\Y\nonumber\\
		&=\hf \int\dsT^a\wedge(\ce_{abc}\d_\e\W^{bc}+4\ii m\e\g_a\Y)~,
		\label{3.19}
	\end{align}
	where we have denoted  
	\begin{equation}\label{deSitter}
		S_{\rm AdS} =S_{\text{SG}}+S_{\text{super-cosm}}~.
	\end{equation}
This action
is invariant 
under the deformed local supersymmetry transformations (\ref{comptrans}), (\ref{connvar}), (\ref{comptranscosmo}), (\ref{elemtranscosmo}) and (\ref{elemtranscosmo2}), 
provided 
\bea
 \d_\e\W^{bc}&=&2\ii m\ce^{abc}\e\g_a\Y~. \label{3.20b}
 \eea
In the unitary gauge \eqref{unitary}, the action \eqref{deSitter} coincides with that proposed by Howe and Tucker to describe AdS supergravity \cite{HT}.

Alternatively, we can deal with a composite connection obtained by imposing the constraint 
(\ref{constraint}),
which makes the variation \eqref{3.19} vanish. 
 In the reminder of this paper, we will work with the condition (\ref{constraint}), which will be necessary for our consideration of (cosmological) topologically massive supergravity theories in section \ref{sect4}.
Requiring the constraint (\ref{constraint}) to be invariant under the transformations 
\begin{subequations}
	\begin{align}\label{2ndSUSY1}
		&\d_\e\Y^\a=-\cD\e^\a-\frac{m}{2}(\e\g_a)^\a E^a~,\\
		&\d_\e E^a=2\ii\e\g^a\Y\label{2ndSUSY2}~,
	\end{align}
\end{subequations}
we can determine a non-trivial variation of the connection. In particular, one finds
\begin{equation}
	E^b\wedge\d_\e\W^a\,_b=-2\ii\e\g^a\cD\Y-\ii m\e E^b\wedge\g_b\g^a\Y~,
\end{equation}
which has the unique solution for the dual connection $\W_{ma}:=\frac{1}{2}\ce_{abc}\W_m\,^{bc}$,
\begin{equation}\label{2ndSUSY3}
	\d_\e\W_{ma}=-2\ii\e\left(\g_m\mfF_a-\frac{1}{2}E_{ma}\g_b\mfF^b\right)+\ii m\e\left(\ce_{abc}E_m\,^b\Y^c-\g_a\Y_m\right)~,
\end{equation}
where 
	\bea\label{dualgrav}
	\hodge\cD\Y=\dd x^m\mfF_m~, \qquad \mfF_m:=\frac{1}{2}\ce_{mnp}\mfF^{np}
	\eea
is the Hodge dual of the gravitino field strength
\bea\label{gravstrength}
\cD\Y=\frac{1}{2}\dd x^m\wedge\dd x^n\mfF_{nm}~, 
\qquad \mfF_{nm}:=\cD_n\Y_m-\cD_m\Y_n=-\mfF_{mn}~. 
\eea 
When $m=0$, this transformation law is compatible with $\d_\e\W_{ma}=0$ since this variation vanishes when $\Y$ is on-shell, $\cD\Y=0$ \cite{DeserKay}.

\section{Topologically massive supergravity} \label{sect4}

 A unique feature of three dimensions is the existence of Chern-Simons terms that can be used to define topologically massive couplings
 \cite{Siegel,JT,Schonfeld,DJT1,DJT2}. 
 
 \subsection{Conformal supergravity} 
 
 Here we study a generalisation of the $\cN=1$ supersymmetric Lorentz Chern-Simons action \cite{DeserKay} which involves the Goldstone fields 
 $X^a $ and $\Q^\a$. We consider the action 
 \begin{equation}\label{Stopo}
	S_{\text{CSG}}=S_{\text{LCS}}+S_{\text{FCS}}~,
\end{equation}
where 
\begin{align}\label{SLCS}
	S_{\text{LCS}}&=\frac{1}{2}\int\tr\left(\W\wedge\dd\W-\frac{1}{3}\W\wedge\W\wedge\W\right)\nonumber\\
	&=\frac{1}{4}\int\dd^3x\,E\,\ce^{mnp}\left(\W_m\,^{ab}R_{npab}+\frac{2}{3}\W_{m\ph{a}b}^{\ph{m}a}\W_{n\ph{b}c}^{\ph{n}b}\W_{p\ph{c}a}^{\ph{p}c}\right)
\end{align}
is the Lorentz Chern-Simons term, and 
 \begin{align}\label{SFCS}
	S_{\text{FCS}}&=2\ii\int\Big(\cD\Y^\a\wedge\hodge\cD\Y_\a+\hodge\cD\Y^\a\wedge E_{\a\b}\wedge\hodge\cD\Y^\b\Big)\nonumber\\
	&=-2\ii\int\dd^3x\,E\,\mfF^a
	\mfF_a
\end{align} 
the fermionic Chern-Simons term. The latter involves the gravitino field strength (\ref{gravstrength}) and its Hodge dual (\ref{dualgrav}).
In the unitary gauge \eqref{unitary}, the functional \eqref{Stopo} coincides with the $\cN=1$ supersymmetric Lorentz Chern-Simons action \cite{DeserKay} which is also known as the action for $\cN=1$ conformal supergravity \cite{vN}.

 We endeavour to demonstrate that the action \eqref{Stopo}
is invariant under the local supersymmetry transformations (\ref{2ndSUSY1}), (\ref{2ndSUSY2}) and (\ref{2ndSUSY3}). The elementary fields $\y^\a$ and $e^a$ transform according to (\ref{elemtranscosmo}) and (\ref{elemtranscosmo2}). As before, it is assumed that the $\d_\e\W$-dependence in these transformation laws is such that the composite fields $\Y^\a$ and $E^a$ remain unchanged when the connection is perturbed $\W\rightarrow\W+\d_\e\W$. 

Once this has been achieved, we can couple the action (\ref{Stopo}) to the AdS supergravity action (\ref{deSitter}) giving a generalisation of cosmological topologically massive supergravity proposed in \cite{Deser}.
However, we first consider topologically massive supergravity without a cosmological term 
\cite{DeserKay} by restricting to the case $m=0$.

Let us  compute variations of the action (\ref{Stopo}) in parts, beginning with the variation of the Lorentz Chern-Simons term (\ref{SLCS}),
\begin{equation}\label{SLCSvar}
	\d S_{\text{LCS}}=\ii\int\dd^3x\,ER(\e\g_a\mfF^a)+4\ii\int\dd^3x\,E\,G^{ab}(\e\g_b\mfF_a)~,
\end{equation}
where $G_{ab}=R_{ab}-\frac{1}{2}\h_{ab}R$ is the Einstein tensor, $R_{ab}=R^{c}{}_{a{c}b}$ is the Ricci tensor and $R=-2\h^{ab} G_{ab}=\h^{ab} R_{ab}$ is the Ricci scalar. Varying the fermionic Chern-Simons term (\ref{SFCS}) with respect to $\Y$ gives
\begin{equation}\label{SFCSPsivar}
	\d_\Y S_{\text{FCS}}=-2\ii\int\dd^3x\,E\Big(G^{ab}(\e\g_a\mfF_b)+\ce^{abc}G_{ba}(\e \mfF_c)+G^{ab}(\e\g_b\mfF_a)+\frac{1}{2}R(\e\g_a\mfF^a)\Big)~,
\end{equation}
where we have used the second relation in \eqref{3.66}.
Combining the variations (\ref{SLCSvar}) and (\ref{SFCSPsivar}) results in the cancellation of the Ricci scalar curvature terms leaving
\begin{equation}\label{combinedvar}
	\d S_{\text{LCS}}+\d_\Y S_{\text{FCS}}=2\ii\int\dd^3x\,E\left\{G^{ab}(\e\g_b\mfF_a)-G^{ab}(\e\g_a\mfF_b)-\ce^{abc}G_{ba}(\e \mfF_c)\right\}~.
\end{equation}
Let us introduce the Hodge dual of the antisymmetric part $R_{[ab]}$ of the Ricci tensor 
\begin{equation}
	\hodge R^a=\frac{1}{2}\ce^{abc}R_{bc}~, \label{4.8}
\end{equation}
so that the combined variation (\ref{combinedvar}) takes the form
\begin{equation}
	\d S_{\text{LCS}}+\d_\Y S_{\text{FCS}}=2\ii\int\dd^3x\,E\left\{G^{ab}(\e\g_b\mfF_a)-G^{ab}(\e\g_a\mfF_b)+2\hodge R^a(\e \mfF_a)\right\}~.
\end{equation}
With some algebraic manipulations, this combination can be brought to the simplified form of a single term involving \eqref{4.8}
\begin{equation}
	\d S_{\text{LCS}}+\d_\Y S_{\text{FCS}}=4\ii\int\dd^3x\,E\,\hodge R^a(\e\g_b\g_a\mfF^b)~,
\end{equation}
which upon inserting the relation (\ref{ModBianchi}) and substituting the identity (\ref{prod}) becomes
\begin{equation}\label{SLCSPsiSFCSvar}
	\d S_{\text{LCS}}+\d_\Y S_{\text{FCS}}=2\int\dd^3x\,E\left\{2(\mfF^b\g^a\Y_b)(\e \mfF_a)+2\ce^{abc}(\mfF_d\g_b\Y^d)(\e\g_c\mfF_a)\right\}~.
\end{equation}

Next we vary the action (\ref{SFCS}) with respect to the composite field $E^a$. This variation reads
\begin{align}\label{ESFCSvar}
	\d_ES_{\text{FCS}}=&\,2\int\dd^3x\,E\big\{-2(\mfF^a\mfF_a)(\e\g^b\Y_b)-2\ce^{abc}(\mfF_b\g_c\mfF_a)(\e\g_d\Y^d)\nonumber\\
	&+4(\mfF^a\mfF_b)(\e\g^b\Y_a)+4\ce^{abc}(\mfF^d\g_c\mfF_a)(\e\g_b\Y_d)\big\}~.
\end{align}

The final variation to be computed is the variation of $S_{\text{FCS}}$ (\ref{SFCS}) with respect to the Lorentz connection. Direct calculations give 
\begin{align}\label{ConnSFCSvar}
	\d_\W S_{\text{FCS}}=&\,2\int\dd^3x\,E\big\{2\ce^{abc}(\mfF^d\g_d\Y_c)(\e\g_b\mfF_a)-2\ce^{abc}(\mfF_a\g_d\Y_c)(\e\g_b\mfF^d)\nonumber\\
	&-2\ce^{abc}(\mfF^d\g_a\Y_c)(\e\g_b\mfF_d)+2(\mfF^a\Y_b)(\e\g_a\mfF^b)\big\}~.
\end{align}

In order to show that the total variation of the action (\ref{Stopo}) vanishes, we will need to perform systematic Fierz rearrangements on the individual terms contained within the variations of (\ref{SLCSPsiSFCSvar}) and (\ref{ConnSFCSvar}) such that all terms have products of the form $(\mfF\mfF)(\e\Y)$, potentially with gamma matrices wedged between the fields. Note that the variation (\ref{ESFCSvar}) is already in the desired form and so will not require a Fierz rearrangement of its terms. After a series of tedious calculations guided by the use of the Fierz rearrangement rule for two-component spinors (\ref{Fierz}), we achieve the desired forms of the variations (\ref{SLCSPsiSFCSvar}) and (\ref{ConnSFCSvar}):
\begin{subequations}
	\begin{align}\label{Fierzvar1}
		\d S_{\text{LCS}}+\d_\Y S_{\text{FCS}}=&\,2\int\dd^3x\,E\big\{-3(\mfF^a\mfF^b)(\e\g_b\Y_a)+(\mfF_b\g^a\mfF_a)(\e\Y^b)\nonumber\\
		&-\ce^{abc}(\mfF^d\g_c\mfF_a)(\e\g_b\Y_d)\big\}~,
	\end{align}
	\begin{align}\label{Fierzvar2}
		\d_\W S_{\text{FCS}}=&\,2\int\dd^3x\,E\big\{(\mfF^a\g_a\mfF^b)(\e\Y_b)-(\mfF^a\mfF^b)(\e\g_a\Y_b)+2(\mfF^a\mfF_a)(\e\g^b\Y_b)\nonumber\\
		&+2\ce^{abc}(\mfF_a\g_b\mfF^d)(\e\g_d\Y_c)+2\ce^{abc}(\mfF_b\g_d\mfF_a)(\e\g^d\Y_c)\nonumber\\
		&+2\ce^{abc}(\mfF_a\g_d\mfF^d)(\e\g_b\Y_c)-\ce^{abc}(\mfF_b\g_c\mfF^d)(\e\g_a\Y_d)\big\}~.
	\end{align}
\end{subequations}
Summing all variations (\ref{Fierzvar1}), (\ref{Fierzvar2}) and (\ref{ESFCSvar}) gives
\begin{align}
	\d S_{\text{CSG}}&=\d S_{\text{LCS}}+\d_\Y S_{\text{FCS}}+\d_ES_{\text{FCS}}+	\d_\W S_{\text{FCS}}\nonumber\\
	&=4\int\dd^3x\,E\Big\{\ce^{abc}(\mfF^d\g_c\mfF_a)(\e\g_b\Y_d)-\ce^{abc}(\mfF_b\g_c\mfF_a)(\e\g_d\Y^d)\nonumber\\
	&\quad+\ce^{abc}(\mfF_a\g_d\mfF^d)(\e\g_b\Y_c)+\ce^{abc}(\mfF_b\g_d\mfF_a)(\e\g^d\Y_c)\nonumber\\
	&\quad+\ce^{abc}(\mfF_a\g_b\mfF^d)(\e\g_d\Y_c)\Big\}~.
\end{align}
The combination in curly brackets can be rewritten in the equivalent form
\begin{align}\label{rewritten}
	&\frac{1}{2}(\e\g_b\Y_d)\Big[\ce^{bac}\big\{(\mfF_a\g_c\mfF^d)+(\mfF^d\g_a\mfF_c)+(\mfF_c\g^d\mfF_a)\big\}\nonumber\\
	&\qquad\qquad+\ce^{dac}\big\{(\mfF_a\g_c\mfF^b)+(\mfF^b\g_a\mfF_c)+(\mfF_c\g^b\mfF_a)\big\}\Big]\nonumber\\
	-&\ce_{abc}(\e\g_d\Y^d)(\mfF^b\g^c\mfF^a)+\frac{1}{2}\ce^{bac}(\e\g_b\Y_d)(\mfF_a\g_c\mfF^d)-\frac{1}{2}\ce^{bac}(\e\g_b\Y_d)(\mfF^d\g_a\mfF_c)\nonumber\\
	-&\frac{1}{2}\ce_{cad}(\e\g^b\Y^d)(\mfF^c\g_b\mfF^a)-\frac{1}{2}\ce^{bac}(\e\g_b\Y_d)(\mfF_c\g^d\mfF_a)+\frac{1}{2}\ce^{dac}(\e\g_b\Y_d)(\mfF_a\g_c\mfF^b)\nonumber\\
	-&\frac{1}{2}\ce^{dac}(\e\g_b\Y_d)(\mfF^b\g_a\mfF_c)-\ce_{abd}(\e\g^b\Y^d)(\mfF^c\g_c\mfF^a)~.
\end{align}
Now if we consider the first cycled combination in curly brackets
\begin{equation}
	\mfF_a\g_c\mfF_d+\mfF_d\g_a\mfF_c+\mfF_c\g_d\mfF_a\equiv \cX_{acd}~,
\end{equation}
we notice that $\cX_{acd}$ is totally antisymmetric in its indices and therefore
\begin{equation}\label{Xmpq}
	\cX_{acd}=k\ce_{acd}=-\frac{1}{2}\ce^{bef}(\mfF_b\g_e\mfF_f)\ce_{acd}~.
\end{equation}
Similarly, applying the same trick for the second cycled term in curly brackets leads to the same result. As a consequence of these observations, we cancel the term in (\ref{rewritten}) proportional to $(\e\g_d\Y^d)$. After additional cancellations within the combination (\ref{rewritten}), it reduces to three remaining terms,
\begin{equation}
	-\frac{1}{2}\ce_{cad}(\mfF^c\g_b\mfF^a)(\e\g^b\Y^d)-\frac{1}{2}\ce_{bac}(\mfF^c\g^d\mfF^a)(\e\g^b\Y_d)-\ce_{abd}(\mfF^c\g_c\mfF^a)(\e\g^b\Y^d)~.
\end{equation}
This combination may be shown to be identically zero. 

As a result, we have demonstrated that the conformal supergravity action \eqref{Stopo}
is invariant under the local supersymmetry transformations (\ref{2ndSUSY1}), (\ref{2ndSUSY2}) and (\ref{2ndSUSY3}). 
This implies that the action 
\begin{equation}\label{TMS}
	S_{\text{TMSG}}
	=\frac{1}{\k} S_{\text{SG}}+ \frac{1}{\m} S_{\text{CSG}}
\end{equation} 
is also invariant under the second local supersymmetry, with $\k$ and $\m$ being coupling constants.
In the unitary gauge \eqref{unitary}, the functional \eqref{TMS} turns into the action for $\cN=1$ topologically massive supergravity originally constructed in \cite{DeserKay}.

	\subsection{Cosmological topologically massive supergravity} 

We now incorporate the supersymmetric cosmological term (\ref{cosmoaction}) by demonstrating that the additional variations of the action (\ref{Stopo}) arising from the $m$-dependent transformation terms in (\ref{2ndSUSY1}) and (\ref{2ndSUSY3}) keep the action stationary. The first contribution comes from varying the Lorentz Chern-Simons action (\ref{SLCS}),
\begin{equation}\label{mSLCSvar}
	\d^{(m)}S_{\text{LCS}}=2\ii m\int\dd^3x\,E \,G^{ab}(\e\g_a\Y_b)+4m\int\dd^3x\,E\,(\Y_a\g_b\mfF^a)(\e\Y^b)~.
\end{equation}
The variation of the fermionic Chern-Simons action (\ref{SFCS}) resulting from the $m$-dependent transformation term for the field $\Y$ reads
\begin{align}
	\d_\Y^{(m)}S_{\text{FCS}}=&\,2m\int\dd^3x\,E\big\{2(\e \mfF^a)(\Y_a\g^b\Y_b)-\ce^{abc}(\e\g_a\mfF^d)(\Y_c\g_d\Y_b)\nonumber\\
	&+\ce^{abc}(\e\g_d\mfF^d)(\Y_c\g_a\Y_b)-\ce^{abc}(\e\g_d\mfF_a)(\Y_c\g^d\Y_b)\nonumber\\
	&-\ii G^{ab}(\e\g_a\Y_b)\big\}~.
\end{align}
Combining these two variations results in the cancellation of the terms proportional to $G^{ab}$ leaving
\begin{align}\label{mSLCSSFCS}
	\d^{(m)}S_{\text{LCS}}+\d_\Y^{(m)}S_{\text{FCS}}=&\,2m\int\dd^3x\,E\big\{(\Y_a\g^b\Y_b)(\e \mfF^a)-(\Y_a\Y_b)(\e\g^b\mfF^a)\nonumber\\
	&-\ce^{abc}(\e\g_b\mfF^d)(\Y_d\g_c\Y_a)-\ce^{abc}(\e\g_a\mfF^d)(\Y_c\g_d\Y_b)\nonumber\\
	&+\ce^{abc}(\e\g_d\mfF^d)(\Y_c\g_a\Y_b)-\ce^{abc}(\e\g_d\mfF_a)(\Y_c\g^d\Y_b)\big\}~,
\end{align}
where the additional terms have arised from a Fierz rearrangement of the second term in (\ref{mSLCSvar}). Finally, the $m$-dependent transformation term for the Lorentz connection gives us the contribution
\begin{align}\label{mSFCSconnvar}
	\d_\W^{(m)}S_{\text{FCS}}=&\,2m\int\dd^3x\,E\big\{(\mfF_a\g^b\Y_b)(\e\Y^a)+(\mfF^a\g_b\Y_a)(\e\Y^b)-(\mfF^a\Y_a)(\e\g^b\Y_b)\nonumber\\
	&+(\mfF^a\Y_b)(\e\g^b\Y_a)-\ce^{abc}(\mfF_b\Y_a)(\e\Y_c)-\ce^{abc}(\mfF_a\g_d\Y_c)(\e\g^d\Y_b)\nonumber\\
	&+\ce^{abc}(\mfF^d\g_d\Y_c)(\e\g_a\Y_b)-\ce^{abc}(\mfF^d\g_a\Y_c)(\e\g_d\Y_b)\big\}~.
\end{align}
Following the strategy used in the $m=0$ case, we perform systematic Fierz rearrangments on the individual terms contained within the variation (\ref{mSFCSconnvar}) such that all terms have products of the form $(\e \mfF)(\Y\Y)$, potentially with gamma matrices wedged between the fields. After applying the Fierz rearrangement rule (\ref{Fierz}) on all terms, we arrive at the desired form of the variation (\ref{mSFCSconnvar}),
\begin{align}\label{mSFCSconnFierz}
	\d_\W^{(m)}S_{\text{FCS}}=&\,2m\int\dd^3x\,E\big\{(\e\g^b\mfF^a)(\Y_a\Y_b)-(\e \mfF^b)(\Y_b\g^a\Y_a)\nonumber\\
	&-\ce^{abc}(\e\g^d\mfF_b)(\Y_c\g_d\Y_a)-\ce^{abc}(\e\g_c\mfF^d)(\Y_d\g_a\Y_b)\big\}~.
\end{align} 
Summing the variations (\ref{mSLCSSFCS}) and (\ref{mSFCSconnFierz}) gives
\begin{align}
	\d^{(m)} S_{\text{CSG}}&=\d^{(m)} S_{\text{LCS}}+\d_\Y^{(m)} S_{\text{FCS}}+\d_\W^{(m)} S_{\text{FCS}}\nonumber\\
	&=2m\int\dd^3x\,E\Big\{\ce^{abc}(\e\g_d\mfF^d)(\Y_c\g_a\Y_b)-\ce^{abc}(\e\g_a\mfF^d)(\Y_c\g_d\Y_b)\nonumber\\
	&\quad-2\ce^{abc}(\e\g_b\mfF^d)(\Y_d\g_c\Y_a)\Big\}~.
\end{align}
The combination of these  three  terms may be shown to be identically zero, and therefore
\begin{equation}
	\d^{(m)}S_{\text{CSG}}=\d^{(m)} S_{\text{LCS}}+\d_\Y^{(m)} S_{\text{FCS}}+\d_\W^{(m)} S_{\text{FCS}}=0~.
\end{equation}
Finally we arrive at
\begin{equation}
	\d S_{\text{CSG}}=\d S_{\text{LCS}}+\d_\Y S_{\text{FCS}}+\d_ES_{\text{FCS}}+	\d_\W S_{\text{FCS}}=0~.
\end{equation}

We have demonstrated that the action
\begin{equation}\label{CTMS}
	S_{\text{CTMSG}}
	=\frac{1}{\k} \Big(S_{\text{SG}}+S_{\text{super-cosm}} \Big)+ \frac{1}{\m} S_{\text{CSG}}
\end{equation} 
is invariant under the second local supersymmetry given by (\ref{2ndSUSY1}), (\ref{2ndSUSY2}) and (\ref{2ndSUSY3}), with $\k$ and $\m$ coupling constants.
In the unitary gauge \eqref{unitary}, the functional \eqref{CTMS} turns into the action for $\cN=1$ cosmological topologically massive supergravity originally constructed in \cite{Deser}.

\section{Conclusion}

In this paper we have developed a nonlinear realisation approach to (cosmological) topologically massive $\cN=1$ supergravity in three dimensions. In addition to the supergravity multiplet, the action involves the Goldstone fields $X^a$ and $\Q^\a$ which are purely gauge degrees of freedom with respect to the local super-Poincar\'e translations generated by the parameters $(b^a, \eta^\a)$. The action is invariant under two different local supersymmetries. One of them 
 acts on the Goldstino, while the other supersymmetry leaves the Goldstino invariant. The former can be used to gauge away the Goldstino, and then the resulting action coincides with that given in the literature \cite{Deser}.
 
There is a remarkable feature of uniqueness in the proposed approach to $\cN=1$ supergravity. The explicit structure of the first local supersymmetry \eqref{2.188} is uniquely determined by the coset construction under consideration. 
In the case of topologically massive supergravity \eqref{TMS}, the structure of the second local supersymmetry \eqref{2.222} is modelled on the first one, eq. \eqref{2.188}. 
In the case of cosmological topologically massive supergravity, the second supersymmetry is deformed by $m$-dependent contributions. 

The action  for cosmological topologically massive supergravity, eq.  \eqref{CTMS}, involves two different functionals, which are separately  invariant under the two local supersymmetry transformations. The first functional 
$S_{\rm AdS} = S_{\text{SG}}+S_{\text{super-cosm}} $ is 
a combination of four terms with  fixed relative coefficients. 
The second functional $S_{\text{CSG}}$ is a combination of two terms with  fixed relative coefficients. 
Changing at least  one of the relative coefficients breaks explicitly the second local supersymmetry, and then the resulting action describes a model for spontaneously broken $\cN=1$ supergravity.

In principle, our construction, which is a natural application of the ideas pioneered by Volkov and Soroka \cite{VS,VS2}, may be generalised to include more general models for massive $\cN=1$ supergravity constructed in \cite{Andringa:2009yc, BHRST10} and recast in the superspace setting of \cite{Kuzenko:2015jda}. That would require a further deformation of the 
second local supersymmetry transformation. Of course, 
the massive supergravity theories of \cite{Andringa:2009yc, BHRST10, Kuzenko:2015jda} were constructed using off-shell supergravity techniques, and our nonlinear realisation approach to supergravity is not a competitor to the off-shell methods, simply due to the fact that the second local supersymmetry is on-shell.  It is still quite remarkable that the structure of $\cN=1$ Poincar\'e supergravity is uniquely determined by applying the formalism of nonlinear realisations.

The Volkov-Soroka approach was also inspirational for a recent work 
\cite{CDS} in which the minimal massive gravity theory of \cite{Bergshoeff:2014pca} was shown to be a particular case of a more general `minimal massive gravity' arising upon 
spontaneous breaking of a local symmetry in a Chern-Simons gravity based on a Hietarinta or Maxwell algebra. It would be interesting to extend the construction of \cite{CDS} to the supersymmetric case.
\\

\noindent
{\bf Acknowledgements:}\\
We are grateful to Ian McArthur and Dmitri Sorokin for discussions and suggestions.
The work of SMK is supported in part by the Australian 
Research Council, projects DP200101944 and DP230101629.
The work of JS is supported by the Australian Government Research Training Program Scholarship.

\appendix

\section{Nonlinear realisation approach to 4D gravity} \label{Appendix0}

In this appendix we review the nonlinear realisation approach to gravity proposed by Volkov and Soroka \cite{VS,VS2}. It can be formulated in $d$ spacetime dimensions 
by identifying the proper orthochronous Poincar\'e group $\sISO_0(d-1,1)$ with 
 the set of all $(d+1)\times (d+1)$ matrices of the form
\bea
D(\L , b)  &=& \left(
\begin{array}{cc}
\L^a{}_b ~& b^a  \\
0 ~& 1\\  
\end{array}
\right)~, \qquad 
\L =(\L^a{}_b) \in \sSO_0(d-1,1)~, \quad b=(b^a) \in {\mathbb R}^d~.
\eea
However, we prefer to fix $d=4$ and work with $\sISL(2,{\mathbb C}) $, the universal covering group 
of the proper orthochronous Poincar\'e group $\sISO_0(3,1)$.
Any element $g =(M, b ) \in \sISL(2,{\mathbb C}) $ is a $4 \times 4$ 
matrix of the form
\begin{subequations}\label{SP}
\bea
g &=& S( {\mathbbm 1}_2,  b) \,h (M,0)  \equiv S h ~, 
\label{SP1} \\
S( {\mathbbm 1}_2,  b) &:= &
\left(
\begin{array}{c | c }
  \mathbbm{1}_2  ~& ~ 0 ~  \\
\hline
-{\rm i}\,\tilde{ b}  \phantom{\Big|}  ~& ~\mathbbm{1}_2 ~ 
\end{array}
\right)
=
\left(
\begin{array}{r | c }
\d_\a{}^\b  ~& ~ 0  \\
\hline
-{\rm i}\,b^{{\dot \a} \b}  ~& ~\d^{\dot \a}{}_{\dot \b} 
\end{array}
\right) ~, 
\label{SP2}  \\
{ h(M,0)} &:=& 
\left(
\begin{array}{c | c }
 M  ~& ~ 0   \\
\hline
0  ~& ~(M^{-1})^\dagger 
\end{array}
\right) 
= \left(
\begin{array}{c | c}
M_\a{}^\b  ~& ~ 0  \\
\hline
0  ~& ({\bar M}^{-1})_{\dot \b}{}^{\dot \a}  
\end{array}
\right) ~, 
\label{SP3} 
\eea
\end{subequations}
where $ b^{\ad \b} = b^a \big(\tilde{\s}_a)^{\ad \b}$,
$M =(M_\a{}^\b)  \in \sSL(2,{\mathbb C})$.
The tilde notation in \eqref{SP2} reflects the fact that there are two types of relativistic Pauli matrices, $\s_a =\big( (\s_a)_{\a \bd} \big)$ and $\tilde \s_a = \big((\tilde{\s}_a)^{\ad \b}\big)$, see \cite{WB}.
The group element $S( {\mathbbm 1}_2,  b) $ is parametrised by a real 4-vector $b^a$.

Introduce a Goldstone vector field $V^a (x) $ for spacetime translations.
It takes its values in the homogeneous space 
${\mathbb M}^{4} = \sISL(2,{\mathbb C}) / \sSL(2,{\mathbb C}) $
(Minkowski space) according to the rule:
\bea
{\mathfrak S}(V ) &= &
\left(
\begin{array}{c | c }
  \mathbbm{1}_2  ~& ~ 0  \\
\hline
-{\rm i}\,\tilde{ V}  \phantom{\Big|}  ~& ~\mathbbm{1}_2 ~ 
\end{array}
\right) 
~.
\eea
We define gauge Poincar\'e transformations by  
\bea
g(x): V(x) \to V'(x) ~, \qquad g {\mathfrak S}(V) = {\mathfrak S}(V') h~,
\eea
with $g=Sh$.
This is equivalent to the following transformations of the Goldstone field:
\begin{subequations}
\bea
 S({\mathbbm 1}_2, b ) :\qquad 
 \tilde V' &=& \tilde V + \tilde b ~,\\
 h(M, 0): \qquad \tilde{V}' &=&  (M^\dagger)^{-1} \tilde{V} M^{-1}  ~.
\eea
\end{subequations}

Introduce a connection $ {\mathfrak A}  = \rd x^m {\mathfrak A}_m $ taking its values in the Poincar\'e algebra,
\bea
{\mathfrak A}  &:= &
\left(
\begin{array}{c | c }
  \Omega  ~& ~ 0   \\
\hline
-{\rm i}\,\tilde{ e}  \phantom{\Big|}  ~& ~-\Omega^\dagger ~ \\
\end{array}
\right)
=
\left(
\begin{array}{r | c }
\Omega_\a{}^\b  ~& ~ 0  \\
\hline
-{\rm i}\,e^{{\dot \a} \b}  ~& ~ -\bar \Omega^\ad{}_\bd  \\
\end{array}
\right) ~, 
\eea
and possessing the gauge transformation law
\bea
 {\mathfrak A}' = g  {\mathfrak A} g^{-1} + g \rd g^{-1}~.
 \eea
 Here the one-forms $\Omega_\a{}^\b $ and $\bar \Omega^\ad{}_\bd $ are the spinor counterparts of the Lorentz connection $\Omega^{ab} = \rd x^m \Omega_m{}^{ab} = - \Omega^{ba} $ such that 
 \bea
 \Omega_\a{}^\b = \hf (\s_{ab})_\a{}^\b\, \Omega^{ab} ~, \qquad 
\bar  \Omega^\ad{}_\bd =- \hf (\tilde \s_{ab})^\ad{}_\bd \,\Omega^{ab} ~,
 \eea
 and the matrices $\s_{ab} $ and $\tilde{\s}_{ab}$ are defined as
 $\s_{ab} = -\frac 14 (\s_a \tilde \s_b - \s_b \tilde \s_a)$ and 
 $\tilde \s_{ab} = -\frac 14 ( \tilde \s_a \s_b - \tilde\s_b  \s_a)$.\footnote{This definition agrees with \cite{BK} and differs by sign from \cite{WB}.} 
 The Lorentz connection is an independent field. 
 
 Associated with ${\mathfrak S}$ and ${\mathfrak A}$ is the following connection 
\bea
{\mathbb A} := {\mathfrak S}^{-1}{\mathfrak A} {\mathfrak S} 
+ {\mathfrak S}^{-1} \rd {\mathfrak S} 
~, 
\eea
which is characterised by the gauge transformation law
\bea
{\mathbb A}' = h {\mathbb A} h^{-1} + h \,\rd h^{-1}~,
\label{A.8}
\eea 
for an arbitrary gauge parameter $g = Sh$.
This transformation law tells us that $\mathbb A$ is invariant under all gauge 
transformations of the form $g = S({\mathbbm 1}_2,b)$ which describe local spacetime translations.  
The connection ${\mathbb A} $ is the main object of the Volkov-Soroka construction. 
It has the form 
\bea
{\mathbb A}  &:= &
\left(
\begin{array}{c | c }
  \Omega  ~& ~ 0  \\
\hline
-{\rm i}\,\tilde{ E}  \phantom{\Big|}  ~& ~-\Omega^\dagger ~\\
\end{array}
\right)~,\qquad \tilde{ E} := \tilde{ e} + \cD \tilde{ V} ~,
\eea
where  $\cD$ denotes the covariant derivative, 
\bea
\cD \tilde{ V} &=& \rd  \tilde{ V} -\Omega^\dagger  \tilde{ V} -  \tilde{ V} \Omega~.
 \eea
Equation \eqref{A.8} is equivalent to the following gauge transformation laws:
\begin{subequations}
\bea
\Omega' &=&  M \Omega M^{-1}  +M \rd M^{-1} ~,\\
\tilde{E}' &=&  (M^\dagger)^{-1} \tilde{E} M^{-1} ~. 
\label{A.13b}
\eea
\end{subequations}
We see that the  one-form $E^a$ transforms as a four-vector under the gauge Lorentz  group.

Let us consider a local Poincar\'e translation, $S({\mathbbm 1}_2, b)$. It only acts on the Goldstone vector field $V^a$ and the vierbein $e^a$, 
\bea
V'{}^a = V^a + b^a~, \qquad e'{}^a = e^a - \cD b^a~.
\eea
We have two types of gauge transformations with vector parameters, the general coordinate transformations and the local Poincar\'e translations. The latter gauge freedom can be fixed by imposing the condition $V^a =0$, and then we stay only with the general coordinate invariance. However, we prefer to keep $V^a$ intact.

The curvature tensor is given by 
\bea
{\mathbb R}=\rd {\mathbb A} - {\mathbb A} \wedge {\mathbb A}~, \qquad 
{\mathbb R}' = h {\mathbb R} h^{-1}~.
\label{A.15}
\eea
Its explicit form is 
\bea
{\mathbb R}  &:= &
\left(
\begin{array}{c | c }
  R  ~& ~ 0  \\
\hline
-{\rm i}\,\tilde{ \mathbb T}  \phantom{\Big|}  ~& ~-R^\dagger ~ \\
\end{array}
\right)~,
\eea
where $R = (R_\a{}^\b) $ and $R^\dagger = (\bar R^\ad{}_\bd )$
form the Lorentz curvature, and
\bea
\tilde {\mathbb T} = \rd \tilde E - \tilde E \wedge \Omega + \Omega^\dagger \wedge \tilde E 
 = \cD \tilde E \quad \Longleftrightarrow \quad 
 {\mathbb T}^a = \cD E^a
\eea
is the torsion tensor. 

Using the transformation laws \eqref{A.13b} and \eqref{A.15}, one can immediately engineer gauge-invariant functionals, including the Einstein-Hilbert action 
\bea
S_{\rm EH} = \frac 14 \int \ce_{abcd} E^a \wedge E^b \wedge R^{cd} 
 = \hf \int \rd^4 x\, E \,R 
\label{EH}
\eea
and the cosmological term
\bea
S_{\rm cosm} = \frac{1}{24} \int \ce_{abcd} E^a \wedge E^b \wedge E^{c} \wedge E^d=-\int\dd^4x\,E ~.
\eea
In this formulation, the action for gravity with a cosmological term is 
\bea
S= \frac{1}{\k^2} \Big(S_{\rm EH} +\L S_{\rm cosm} \Big) 
=    \frac{1}{2\k^2}  \int \rd^4 x\, E \,\Big( R - 2\L \Big)~,
\eea
with $\L$ the cosmological constant. Varying this action with respect to the connection
leads to the equation of motion ${\mathbb T}^a=0$.

\section{3D notation and conventions}\label{Appendix A}

In this appendix we collect key formulae of the 3D two-component spinor formalism that is described in \cite{KPT-MvU}. The starting point for setting up this 3D spinor formalism is the 4D relativistic Pauli matrices\footnote{In contrast to the 4D notation used in the previous appendix, here it is useful to denote 4D vector indices by underlined Latin letters.}
\begin{equation}
	(\s_{\underline{a}})_{\a\bd}:=(\id_2,\vec{\s})~,\qquad (\tilde{\s}_{\underline{a}})^{\ad\b}:=(\id_2,-\vec{\s})~,\qquad \underline{a}=0,1,2,3~,
\end{equation}
	where $\vec{\s}=(\s_1,\s_2,\s_3)$ are the Pauli matrices. We remove the matrices with space index $\underline{a}=2$ and obtain the 3D gamma-matrices
\begin{subequations}
\begin{align}\label{gammadown}
	(\s_{\underline{a}})_{\a\bd}\quad &\longrightarrow \quad (\g_a)_{\a\b}=(\g_a)_{\b\a}=(\id_2,\s_1,\s_3)~,\\
	(\tilde{\s}_{\underline{a}})^{\ad\b}\quad &\longrightarrow\quad(\g_a)^{\a\b}=(\g_a)^{\b\a}=\ce^{\a\g}\ce^{\b\d}(\g_a)_{\g\d}\label{gammaup},\quad a=0,1,2 ~.
\end{align}
\end{subequations}

The $(\g_a)_{\a\b}$ and $(\g_a)^{\a\b}$ are invariant tensors of the Lorentz group $\sSO_0(2,1)$. They can be used to convert any three-vector 
$V^a$ into symmetric second-rank spinors
\begin{subequations}
\bea
\check{V} &=& (V_{\a\b})~, \qquad V_{\a\b} = V^a(\g_a)_{\a\b}~; \\
\hat{V} &=& (V^{\a\b})~, \qquad V^{\a\b} = V^a(\g_a)^{\a\b}~.
\label{B3.b}
\eea
\end{subequations}
As is known, the invariance properties of $(\g_a)_{\a\b}$ and $(\g_a)^{\a\b}$ follow from 
the isomorphism $\sSO_0(2,1) \cong \sSL(2,\dsR) /{\mathbb Z}_2$ which is defined by associating with a group element $M \in \sSL(2, {\mathbb R})$ the linear transformation on the vector space of symmetric real $2\times 2$ matrices 
$\check{V} $
\bea
\check{V} \to M \check{V} M^{\rm T}~.
\eea

In the 3D case, the spinor indices are lowered  and raised using the $\sSL(2,\dsR)$ invariant spinor metric $\ce=(\ce_{\a\b}) = -(\ce_{\b \a}) $ and its inverse $\ce^{-1}=(\ce^{\a\b}) = - (\ce^{\b\a}) $, which are normalised by $\ce^{12}=-\ce_{12}=1$.
The rules for lowering and raising the spinor indices are:
\begin{equation}
	\y_\a=\ce_{\a\b}\y^\b~, \qquad
		\y^\a=\ce^{\a\b}\y_\b~.
	\end{equation}
By construction, the $\g$-matrices (\ref{gammadown}) and (\ref{gammaup}) are real and symmetric. 

Properties of the 4D relativistic Pauli matrices imply analogous properties of the 3D $\gamma$-matrices. In particular, for the Dirac matrices
\begin{equation}\label{gammamat}
	\g_a:=\left( (\g_a)_\a\,^\b\right) =\ce^{\b\g}(\g_a)_{\a\g}=(-\ii\s_2,\s_3,-\s_1)
\end{equation}
we have the following identities
\begin{subequations}
	\begin{align}
		\g_a\g_b&=\h_{ab}\id_2+\ce_{abc}\g^c\label{prod}\quad \implies \quad 
		\{\g_a,\g_b\}=2\h_{ab}\id_2~,
		\\
		\g_a\g_b\g_c&=\h_{ab}\g_c-\h_{ac}\g_b+\h_{bc}\g_a+\ce_{abc}\id_2~,\\
		(\g^a)^{\a\b}(\g_a)^{\g\d}&=\ce^{\a\g}\ce^{\d\b}+\ce^{\a\d}\ce^{\g\b}~\label{gammacontraction},
	\end{align}
\end{subequations}
where the 3D Minkowski metric is $\h_{ab}=\h^{ab}=\text{diag}(-1,+1,+1)$, and the Levi-Civita tensors $\ce_{abc}$ and $\ce^{abc}$ are normalised by $\ce_{012}=-\ce^{012}=-1$. 

Throughout this paper, contractions of spinor indices are defined as follows:
\begin{subequations}
	\begin{align}
		\f\c&:=\f^\a\c_\a=\c\f,\qquad \f^2:=\f\f,\\
		\f\g_a\c&:=\f^\a(\g_a)_\a\,^\b\c_\b=-\c\g_a\f.
	\end{align}
\end{subequations}
Here $\phi_\a$ and $\chi_\a$ are arbitrary anti-commuting spinors.

The Dirac matrices (\ref{gammamat}) along with the unit matrix, $\G_A:=\{\id_2,\g_a\}$, form a basis in the linear space of $2\times 2$ matrices. If we define the corresponding set with upper indices, $\G^A:=\{\id_2,\g^a\}$, we have the identity
\begin{equation}
	{\rm tr}\,(\G_A\G^B)=2\d_A\,^B~.
\end{equation}
In accordance with this identity, if $M=(M_\a\,^\b)$ and $N=(N_\a\,^\b)$ are $2\times 2$ matrices, then
\begin{subequations}
\begin{align}\label{basis}
	M_\a\,^\b N_\g\,^\d&=\sum_{A}(C^A)_\a\,^\d(\G_A)_\g\,^\b~,\\
	(C^A)_\a\,^\d&=\frac{1}{2}M_\a\,^\b(\G^A)_\b\,^\g N_\g\,^\d~.\label{coeff}
\end{align}
\end{subequations}
Now let $\y_1^\a$, $\y_2^\a$, $\y_3^\a$ and $\y_4^\a$ be arbitrary two-component spinors. Using the equations (\ref{basis}) and (\ref{coeff}) one can show that
\begin{equation}\label{Fierz}
	(\y_1M\y_2)(\y_3N\y_4)=-\frac{1}{2}(\y_1M\G^AN\y_4)(\y_3\G_A\y_2)~,
\end{equation}
which is the Fierz rearrangement rule for two-component spinors. 

The Levi-Civita tensor with lower curved-space indices, $\ce_{mnp}$, is defined by
\begin{equation}
	\ce_{mnp}=E\e_{mnp}=E_m\,^aE_n\,^bE_p\,^c\ce_{abc}~,
\end{equation}
where $E:=\det(E_m\,^a)$ and $\e_{mnp}$ is the Levi-Civita symbol. Its counterpart 
with upper curved-space indices $\ce^{mnp}$ is
\begin{equation}\label{upperLev}
	\ce^{mnp}=E^{-1}\e^{mnp}=E_a\,^mE_b\,^nE_c\,^p\ce^{abc}~.
\end{equation}

In three dimensions, any vector $F^a$ can be equivalently realised as a symmetric second-rank spinor 
$F_{\a\b}= F_{\b\a}$ or as an antisymmetric second-rank tensor $F_{ab} = -F_{ba}$. 
The former realisation is obtained using the gamma-matrices:
\begin{equation}
	F_{\a\b}:=(\g^a)_{\a\b}F_a=F_{\b\a}~,\qquad F^a=-\frac{1}{2}(\g^a)^{\a\b}F_{\a\b}~.
	\label{B.11}
\end{equation}
The antisymmetric tensor $F_{ab}$ is the Hodge-dual of $F_a$,
\begin{equation}
	F_{ab}=-\ce_{abc}F^c~, \qquad F_a=\frac{1}{2}\ce_{abc}F^{bc}~.
	\label{B.12}
\end{equation}
The symmetric spinor $F_{\a\b}$ is defined in terms of $F_{ab}$ as follows
\begin{equation}
	F_{\a\b}
	=\frac{1}{2}(\g^a)_{\a\b}\ce_{abc}F^{bc}~.
	\label{B.13}
\end{equation}
We emphasise that the three algebraic objects $F_a$, $F_{ab}$ and $F_{\a \b}$ 
are equivalent  to each other. The corresponding inner products are related to each other as follows:
\bea
-F^aG_a=
\hf F^{ab}G_{ab}=\hf F^{\a\b}G_{\a\b}
~.
\eea
More details can be found in \cite{KLT-M11}.

\section{The first Bianchi identity} \label{appendixB}

The first Bianchi identity  is given by
\begin{equation}
	\cD\cD E^a=E^b\wedge R^a\,_b~.
\end{equation}
Requiring the supersymmetric torsion to vanish gives
\begin{align}
	&\dsT^a=\cD E^a-\ii\Y\wedge\g^a\Y=0 \quad
	\Longrightarrow\quad \cD\cD E^a=2\ii\Y\wedge\g^a\cD\Y~.
\end{align}
Therefore,
\begin{equation}
	E^b\wedge R^a\,_b=2\ii\Y\wedge\g^a\cD\Y~.
\end{equation}
By expanding this expression into its components and contracting with $\ce^{mpn}$ we obtain
\begin{align}\label{Bianchi}
	\ce^{mpn}R_{mpn}\,^a&=-4\ii\ce^{mpn}(\Y_m\g^a\cD_p\Y_n)
	=-2\ii\ce^{mpn}\big(\Y_m\g^a(\cD_p\Y_n-\cD_n\Y_p)\big)\nonumber\\
	&=-2\ii\ce^{mpn}(\Y_m\g^a\mfF_{pn})
	=-4\ii(\Y_m\g^a\mfF^m)~. 
\end{align}
The relation (\ref{Bianchi}) implies that the dual of the antisymmetric part $R_{[ab]}$ of the 
Ricci tensor, eq. \eqref{4.8},  can be expressed in terms of the fermionic fields:
\begin{equation}\label{ModBianchi}
	\hodge R^a=\ii\Y_b\g^a\mfF^b~.
\end{equation}


\begin{footnotesize}
	
\end{footnotesize}


\begin{thebibliography}{66}
		
\bibitem{CWZ}
  S.~R.~Coleman, J.~Wess and B.~Zumino,
  ``Structure of phenomenological Lagrangians. 1,''
  Phys.\ Rev.\  {\bf 177}, 2239 (1969).
  
  \bibitem{CCWZ}
  C.~G.~Callan Jr., S.~R.~Coleman, J.~Wess and B.~Zumino,
  ``Structure of phenomenological Lagrangians. 2,''
  Phys.\ Rev.\  {\bf 177},  2247 (1969).
  
\bibitem{Isham}
C.~J.~Isham,
``A group-theoretic approach to chiral transformations,''
  Nuovo Cim.\  A {\bf 59},  356 (1969).

\bibitem{SalamS1}
A.~Salam and J.~A.~Strathdee,
``Nonlinear realizations. 1. The Role of Goldstone bosons,''
Phys. Rev. \textbf{184}, 1750 (1969).

\bibitem{Volkov}
D.~V.~Volkov, ``Phenomenological Lagrangians,''
Sov. J. Particles Nucl. {\bf 4}, 1 (1973).

\bibitem{Ogievetsky}
V.~I.~Ogievetsky, ``Nonlinear realizations of internal and space-time symmetries,''
in {\it Proceeding of 10th Karpacz Winter School of Theoretical Physics, Vol. 1},
Wroslaw, 1974, pp. 117--132.

\bibitem{SalamS2}
A.~Salam and J.~A.~Strathdee,
``Nonlinear realizations. 2. Conformal symmetry,''
Phys. Rev. \textbf{184}, 1760 (1969).

\bibitem{ISS1}
C.~J.~Isham, A.~Salam and J.~A.~Strathdee,
``Spontaneous breakdown of conformal symmetry,''
Phys. Lett. B \textbf{31}, 300 (1970).

\bibitem{ISS2}
C.~J.~Isham, A.~Salam and J.~A.~Strathdee,
``Nonlinear realizations of space-time symmetries. Scalar and tensor gravity,''
Annals Phys. \textbf{62}, 98 (1971).

\bibitem{Zumino} B. Zumino, 
``Effective Lagrangians and broken symmetries," 
in {\it Lectures on Elementary Particles and Quantum Field Theory,
Vol. 2}, S. Deser, M. Grisaru and H. Pendleton (Eds.),
Cambridge, Mass. 1970, pp. 437-500.

\bibitem{IO}
E.~A.~Ivanov and V.~I.~Ogievetsky,
``The inverse Higgs phenomenon in nonlinear realizations,'' Theor. Math. Phys. 
{\bf 25}, 1050 (1975) 
[Teor. Mat. Fiz. \textbf{25}, 164 (1975)].

\bibitem{McArthur}
I.~N.~McArthur,
``Nonlinear realizations of symmetries and unphysical Goldstone bosons,''
JHEP \textbf{11}, 140 (2010)
[arXiv:1009.3696 [hep-th]].
	
\bibitem{VS} 
  D.~V.~Volkov and V.~A.~Soroka,
  ``Higgs effect for Goldstone particles with spin 1/2,''
  JETP Lett.\  {\bf 18}, 312 (1973)
  [Pisma Zh.\ Eksp.\ Teor.\ Fiz.\  {\bf 18}, 529 (1973)].

\bibitem{VS2}
D.~V. Volkov and V.~A. Soroka, ``Gauge fields for symmetry group with
  spinor parameters,''   Theor. Math. Phys. {\bf 20}, 829 (1974)  
[Teor. Mat. Fiz. {\bf 20}, 291(1974)].

\bibitem{SW2}
K.~S.~Stelle and P.~C.~West,
``Spontaneously broken de Sitter symmetry and the gravitational holonomy group,''
Phys. Rev. D \textbf{21}, 1466 (1980).

\bibitem{IvanovN}
E.~A.~Ivanov and J.~Niederle,
``Gauge formulation of gravitation theories. 1. The Poincar\'e, de Sitter and conformal cases,''
Phys. Rev. D \textbf{25}, 976 (1982).

\bibitem{Utiyama}
R.~Utiyama,
``Invariant theoretical interpretation of interaction,''
Phys. Rev. \textbf{101}, 1597 (1956).

\bibitem{Kibble}
T.~W.~B.~Kibble,
``Lorentz invariance and the gravitational field,''
J. Math. Phys. \textbf{2}, 212 (1961).

\bibitem{Sciama} D. W. Sciama, ``The analogy between charge and spin in general relativity,'' in {\it Recent Developments in General Relativity},  Pergamon, Oxford, 1962,
pp. 415--438. (Reprinted in \cite{Blagojevic:2013xpa}.)

\bibitem{Hehl}
F.~W.~Hehl, P.~Von Der Heyde, G.~D.~Kerlick and J.~M.~Nester,
``General relativity with spin and torsion: Foundations and prospects,''
Rev. Mod. Phys. \textbf{48}, 393 (1976).

\bibitem{Blagojevic:2013xpa}
M.~Blagojevi\'c and F.~W.~Hehl,
{\it Gauge Theories of Gravitation: A Reader with Commentaries},
Imperial College Press, London, 2013.

\bibitem{Volkov2}
D.~V.~Volkov,
``Supergravity before 1976,'' in: 
{\it History of Original Ideas and Basic Discoveries in Particle Physics}, 
H. B. Newman and T. Ypsilantis (Eds.), 
Plenum Press, New York (1996), pp. 663-675
[arXiv:hep-th/9410024 [hep-th]].



\bibitem{BMST} 
I.~Bandos, L.~Martucci, D.~Sorokin and M.~Tonin,
``Brane induced supersymmetry breaking and de Sitter supergravity,''
JHEP {\bf 1602}, 080 (2016)
[arXiv:1511.03024 [hep-th]].



\bibitem{VA}
D.~V.~Volkov and V.~P.~Akulov,
``Possible universal neutrino interaction,''
  {JETP Lett.\  {\bf 16}, 438 (1972)}   
  [Pisma Zh.\ Eksp.\ Teor.\ Fiz.\   {\bf 16},  621 (1972)]; 
  ``Is the neutrino a Goldstone particle?,''
  Phys.\ Lett.\  B {\bf 46}, 109 (1973).

\bibitem{AV}
V.~P. Akulov and D.~V. Volkov, ``Goldstone fields with spin 1/2,''
   Theor. Math. Phys. {\bf 18}, 28 (1974)  28 [Teor. Mat. Fiz. {\bf 18}, 39 (1974)].


\bibitem{K2021}
S.~M.~Kuzenko,
``Local supersymmetry: Variations on a theme by Volkov and Soroka,''
Proc. Roy. Soc. Lond. A \textbf{479}, 20230022 (2023)
[arXiv:2110.12835 [hep-th]].

\bibitem{DZ}
S.~Deser and B.~Zumino,
``Consistent supergravity,''
Phys. Lett. B \textbf{62}, 335 (1976).

\bibitem{FvNF}
D.~Z.~Freedman, P.~van Nieuwenhuizen and S.~Ferrara,
``Progress toward a theory of supergravity,''
Phys. Rev. D \textbf{13}, 3214 (1976).

\bibitem{HT}
P.~S.~Howe and R.~W.~Tucker,
 ``Local supersymmetry in (2+1) dimensions.
1. Supergravity and differential forms,''
J.\ Math.\ Phys.\ {\bf 19}, 869 (1978).

\bibitem{HT2} 
P.~S.~Howe and R.~W.~Tucker,
``A locally supersymmetric and reparametrization invariant action
for a spinning membrane,'' 
J. Phys.\ A {\bf 10}, L155 (1977); 
 ``Local supersymmetry in
(2+1) dimensions. 2. An action for a spinning membrane,''
J.\ Math.\ Phys.\ {\bf 19}, 981 (1978).

\bibitem{DeserKay}
S.~Deser and J.~H.~Kay
``Topologically massive supergravity,''
Phys.\ Lett.\ B {\bf 120}, 97 (1983).

\bibitem{Deser}
S.~Deser,
``Cosmological topological supergravity,''
 in {\it Quantum Theory Of Gravity}, S. M. Christensen (Ed.), 
 Adam Hilger, Bristol, 1984, pp. 374-381.
%



\bibitem{BG}
M.~Brown and S.~J.~Gates Jr.,
``Superspace Bianchi identities and the supercovariant derivative,'' 
Annals Phys.\ {\bf 122}, 443 (1979).

\bibitem{GGRS}
S.~J.~Gates Jr., M.~T.~Grisaru, M.~Rocek and W.~Siegel,
{\emph{Superspace, or One Thousand and One
		Lessons in Supersymmetry}},
Front.\ Phys.\ {\bf 58}, 1 (1983)
\href{https://arxiv.org/abs/hep-th/0108200}{[arXiv:hep-th/0108200]}.



\bibitem{ZP88}
  B.~M.~Zupnik and D.~G.~Pak,
   ``Superfield formulation of the simplest three-dimensional gauge theories and
  conformal supergravities,''
  Theor.\ Math.\ Phys.\  {\bf 77} (1988) 1070
  [Teor.\ Mat.\ Fiz.\  {\bf 77} (1988) 97].
  
\bibitem{ZP89} 
B.~M.~Zupnik and D.~G.~Pak,
``Differential and integral forms in supergauge theories and supergravity,''
Class.\ Quant.\ Grav.\  {\bf 6}, 723 (1989).
  
\bibitem{LR89}
  U.~Lindstr\"om and M.~Ro\v{c}ek,
  ``Superconformal gravity in three dimensions as a gauge theory,''
  Phys.\ Rev.\ Lett.\  {\bf 62}, 2905 (1989).
  


\bibitem{KLT-M11}
S.~M.~Kuzenko, U.~Lindstr\"om and G.~Tartaglino-Mazzucchelli,
  ``Off-shell supergravity-matter couplings in three dimensions,''
  JHEP {\bf 1103}, 120 (2011)
  [arXiv:1101.4013 [hep-th]].

  
  
\bibitem{vN}
P.~van Nieuwenhuizen,
``D = 3 conformal supergravity and Chern-Simons terms,''
Phys.\ Rev.\  D {\bf 32}, 872 (1985).

\bibitem{Uematsu}
  T.~Uematsu,
  ``Structure of N=1 conformal and Poincare supergravity in (1+1)-dimensions
  and (2+1)-dimensions,''
  Z.\ Phys.\  C {\bf 29}, 143 (1985);
``Constraints and actions in two-dimensional and three-dimensional N=1
conformal supergravity,''
Z.\ Phys.\  C {\bf 32}, 33 (1986).

\bibitem{KT-M12} 
 S.~M.~Kuzenko and G.~Tartaglino-Mazzucchelli,
 ``Conformal supergravities as Chern-Simons theories revisited,''
  JHEP {\bf 1303}, 113 (2013)
  [arXiv:1212.6852 [hep-th]].


\bibitem{BKNT-M1} 
  D.~Butter, S.~M.~Kuzenko, J.~Novak and G.~Tartaglino-Mazzucchelli,
  ``Conformal supergravity in three dimensions: New off-shell formulation,''
  JHEP {\bf 1309}, 072 (2013)
  [arXiv:1305.3132 [hep-th]].
 
\bibitem{BKNT-M2}
D.~Butter, S.~M.~Kuzenko, J.~Novak and G.~Tartaglino-Mazzucchelli,
 ``Conformal supergravity in three dimensions: Off-shell actions,''
  JHEP {\bf 1310}, 073 (2013)
  [arXiv:1306.1205 [hep-th]].
  
  

\bibitem{AT}
  A.~Ach\'ucarro and P.~K.~Townsend,
  ``A Chern-Simons action for three-dimensional anti-de Sitter supergravity
 theories,''
  Phys.\ Lett.\  B {\bf 180}, 89 (1986).

\bibitem{Dereli:1977yx}
T.~Dereli and S.~Deser,
``Fermionic Goldstone-Higgs effect in (2+1)-dimensional supergravity,''
J. Phys. A \textbf{11}, L27 (1978).

\bibitem{Hohm:2012vh}
O.~Hohm, A.~Routh, P.~K.~Townsend and B.~Zhang,
``On the Hamiltonian form of 3D massive gravity,''
Phys. Rev. D \textbf{86}, 084035 (2012)
[arXiv:1208.0038 [hep-th]].

\bibitem{Routh}
A.~Routh,
``Hamiltonian form of topologically massive supergravity,''
Phys. Rev. D \textbf{88}, no.2, 024022 (2013)
[arXiv:1301.7671 [hep-th]].

\bibitem{KPT-MvU} 
S.~M.~Kuzenko, J.~Park, G.~Tartaglino-Mazzucchelli and R.~von Unge
``Off-shell superconformal nonlinear sigma-models in
three dimensions,''
JHEP {\bf 1101}, 146 (2010)
\href{https://arxiv.org/abs/1011.5727}{[arXiv:hep-th/1011.5727]}.
		
		
		
\bibitem{TvN}
P.~K.~Townsend and P.~van Nieuwenhuizen,
``Geometrical interpretation of extended supergravity,''
Phys. Lett. B \textbf{67}, 439 (1977).

\bibitem{CW}
A.~H.~Chamseddine and P.~C.~West,
``Supergravity as a gauge theory of supersymmetry,''
Nucl. Phys. B \textbf{129}, 39 (1977).


\bibitem{Siegel}
W.~Siegel,  ``Unextended superfields in extended supersymmetry,''
Nucl.\ Phys.\  B {\bf 156}, 135 (1979).

\bibitem{JT} 
  R.~Jackiw and S.~Templeton,
  ``How super-renormalizable interactions cure their infrared divergences,''
  Phys.\ Rev.\ D {\bf 23}, 2291 (1981).

\bibitem{Schonfeld} 
J.~F.~Schonfeld, ``A mass term for three-dimensional gauge fields,''
  Nucl.\ Phys.\ B {\bf 185}, 157 (1981).

\bibitem{DJT1}
  S.~Deser, R.~Jackiw and S.~Templeton,
  ``Three-dimensional massive gauge theories,''
  Phys.\ Rev.\ Lett.\  {\bf 48}, 975 (1982).

\bibitem{DJT2}
  S.~Deser, R.~Jackiw and S.~Templeton,
  ``Topologically massive gauge theories,''
  Annals Phys.\  {\bf 140}, 372 (1982)
  [Erratum-ibid.\  {\bf 185}, 406 (1988)].
 

 \bibitem{WB} J.~Wess and J.~Bagger,
{\it Supersymmetry and Supergravity},
Princeton University Press, Princeton, 1992.


\bibitem{BK} I.~L.~Buchbinder and S.~M.~Kuzenko,
{\it Ideas and Methods of Supersymmetry and
Supergravity or a Walk Through Superspace},
IOP, Bristol, 1998.



\bibitem{Andringa:2009yc} 
  R.~Andringa, E.~A.~Bergshoeff, M.~de Roo, O.~Hohm, E.~Sezgin and P.~K.~Townsend,
  ``Massive 3D supergravity,''
  Class.\ Quant.\ Grav.\  {\bf 27}, 025010 (2010)
  [arXiv:0907.4658 [hep-th]].

\bibitem{BHRST10} 
  E.~A.~Bergshoeff, O.~Hohm, J.~Rosseel, E.~Sezgin and P.~K.~Townsend,
  ``More on massive 3D supergravity,''
  Class.\ Quant.\ Grav.\  {\bf 28}, 015002 (2011)
  [arXiv:1005.3952 [hep-th]].
  
\bibitem{Kuzenko:2015jda}
S.~M.~Kuzenko, J.~Novak and G.~Tartaglino-Mazzucchelli,
``Higher derivative couplings and massive supergravity in three dimensions,''
JHEP \textbf{09}, 081 (2015)
[arXiv:1506.09063 [hep-th]].

\bibitem{CDS}
D.~Chernyavsky, N.~S.~Deger and D.~Sorokin,
``Spontaneously broken $3d$ Hietarinta/Maxwell Chern\textendash{}Simons theory and minimal massive gravity,''
Eur. Phys. J. C \textbf{80}, no.6, 556 (2020)
[arXiv:2002.07592 [hep-th]].

\bibitem{Bergshoeff:2014pca}
E.~Bergshoeff, O.~Hohm, W.~Merbis, A.~J.~Routh and P.~K.~Townsend,
``Minimal massive 3D gravity,''
Class. Quant. Grav. \textbf{31}, 145008 (2014)
[arXiv:1404.2867 [hep-th]].

%
%
		
\end{thebibliography}
\end{document}